\documentclass[12pt,a4paper]{article}

\usepackage[utf8]{inputenc}
\usepackage[T1]{fontenc}
\usepackage{lmodern}
\usepackage[english]{babel}

\usepackage{amsmath}
\usepackage{siunitx}
\usepackage{pgfplots}

\usepackage{graphicx}
\usepackage{booktabs}
\usepackage{array}
\usepackage{subcaption}
\usepackage{multirow} 
\usepackage{enumitem} 
\usepackage[svgnames]{xcolor}
\usepackage{geometry}
\usepackage{hyperref}
\usepackage{titlesec}
\usepackage{float}
\usepackage{verbatim}
\usepackage{siunitx}

\titleformat*{\section}{\large\bfseries}      
\titleformat*{\subsection}{\large\bfseries}   
\titleformat*{\subsubsection}{\normalsize\bfseries} 

\makeatletter
\renewcommand{\@maketitle}{
    \begin{center}
        {\LARGE \bfseries \@title \par}
        \vskip 1.5em
        {\large \@author \par}
        \vskip 1em
        {\normalsize \@date \par}
    \end{center}
    \vskip 2em
}
\makeatother

\title{\textbf{Vortex-controlled heat transfer in
eight-row plate–fin–and–tube exchangers:
CFD-derived air-side Nusselt correlations} \\
    \vspace{0.5cm}}

\author{
    Mateusz Marcinkowski, Kacper Korab
    \\
    \vspace{0.3cm}
    \small{Faculty of Environmental and Energy Engineering, Cracow University of Technology, Warszawska 24, 31-155 Kraków} \\
    \small{*Corresponding author: mateusz.marcinkowski@pk.edu.pl}
}

\pgfplotsset{compat=1.18}

\begin{document}
\maketitle

\section*{\small Keywords}
compact tube heat exchanger; row by row Nusselt number correlation; vortex shedding; wake interference; RANS CFD;

\section{Abstract}

This paper investigates and characterizes the row-dependent thermal performance of an 8-row finned tube heat exchanger (HEX) with staggered tube arrangement and continuous flat fins. Computational fluid dynamics (CFD) simulations reveal distinct hydraulic–thermal behavior across tube rows, with the first, second, and eighth rows exhibiting noticeably greater heat transfer efficiency than the HEX-wide average under specific flow conditions. \textbf{At} air velocities below \SI{3}{\meter\per\second}, the initial rows demonstrate increased performance, while the eighth row shows comparably enhanced performance at velocities exceeding \SI{6}{\meter\per\second}. Across the fitted range (\SI{2.5}{}–\SI{10}{\meter\per\second}), rows 1, 2, and 8 exceed rows 3–7 by approximately 15–25\% in Nusselt number, indicating that rows 1, 2, and 8 operate under distinct flow–thermal conditions. The study also derives individual Nusselt number correlations for each row through systematic simulations across velocity regimes ranging from \SI{2.5}{} to \SI{10}{\meter\per\second}. The analysis further links these variations to specific flow mechanisms, such as boundary layer development, wake interference, and vortex shedding, which govern the distinct thermal behavior of the first, second, and last rows. These findings could enable dual optimization: reducing capital costs through minimized row count for target thermal output, and lowering operating costs via decreased pressure drop (fan power reduction).

\newpage

\section{Literature Review} 
Finned tube heat exchangers (plate–fin–and–tube designs, FTHE) have been extensively studied due to their widespread use in HVAC, refrigeration, automotive, and power generation. Traditional design approaches rely on average air-side heat transfer coefficients (HTC) across the entire exchanger, neglecting local variations between tube rows. This simplification overlooks important thermal differences that strongly affect exchanger efficiency.

Recent reviews emphasize that most research has focused on overall performance improvements, such as modifying fin geometry, applying vortex generators, or using coatings, while relatively little attention has been paid to detailed row-by-row heat transfer behavior \cite{Sadeghianjahromi2021}. This gap is critical: neglecting local differences may lead to oversizing and inefficient designs, whereas row-specific analysis enables optimization of exchanger depth and fin–tube layout.

Several studies have addressed row-specific heat transfer in FTHE. CFD-based methods \cite{Marcinkowski2022,Marcinkowski2024} demonstrated that the first row typically achieves the highest HTC due to undisturbed inlet flow, while subsequent rows experience reduced performance from wake interference. The last row can recover or even surpass mid-row performance under high Reynolds numbers due to free vortex shedding. Che and Elbel \cite{Che2021} used absorption-based visualization methods to obtain local HTC maps, confirming degradation in inner rows and the influence of fin geometry. Other works investigated design modifications: Zhang et al. \cite{Zhang2021} analyzed convex protrusions on fin surfaces and showed that they regenerate turbulence in wake regions, Hashem-ol-Hosseini et al. \cite{Hashem2020} tested oval tubes with different fin configurations and reported aerodynamic benefits, and Okbaz et al. \cite{Okbaz2020} demonstrated that louvered fins lose effectiveness as row count increases. Kim et al. \cite{Kim2023} studied vibrating fins and observed modest improvements in stagnant flow zones. These works indicate a growing interest in understanding local heat transfer and in methods to influence dead zones and vortex formation regions.

Beyond finned exchangers, studies of bare cylinders and tube bundles provide analogies, since finned tube geometries can be interpreted as tube bundles with lateral confinement imposed by fins. In both systems, flow separation, vortex shedding, and wake interactions govern heat transfer. Zhao et al. \cite{Zhao2020} linked laminar–turbulent transition in tube bundles to non-uniform heat transfer. Zhou et al. \cite{Zhou2021} demonstrated that vortex-induced oscillations in staggered arrays enhance local mixing and Nusselt numbers. Liang et al. \cite{Liang2022} quantified how wake recovery in tube bundles influences downstream heat transfer. Saha et al. \cite{Saha2023} showed that large-scale vortex shedding at high Reynolds numbers governs heat transfer enhancement. Choi et al. \cite{Choi2024} combined PIV and DNS to map vortex–wake interactions and found that coherent shedding structures can increase rear-row HTC under specific spacing ratios.

Taken together, these works show that vortex dynamics, wake interference, and boundary layer disruption explain the uneven distribution of HTC across tube rows. However, most studies have concentrated on shallow exchangers (up to four rows) or on bare-tube bundles without fins. What is still missing is a systematic characterization of vortex-driven heat transfer mechanisms across the full depth of multi-row finned-tube exchangers (e.g., eight rows), together with row-resolved Nusselt correlations. Addressing this gap is the central motivation of the present study.

\section{CFD model: geometry, boundary conditions, and turbulence modelling (concise)}
\label{sec:cfd_concise}

\textit{The full workflow (geometry build, meshing strategy, wall treatment, solver setup, and verification) is detailed in our previous article \cite{Marcinkowski2024}. Here, we summarize only the elements necessary to reproduce the present study with eight rows and further, deeper analysis.}

\subsection{Geometry and computational domain}
\begin{figure}[ht]
    \centering
    \includegraphics[width=0.9\linewidth]{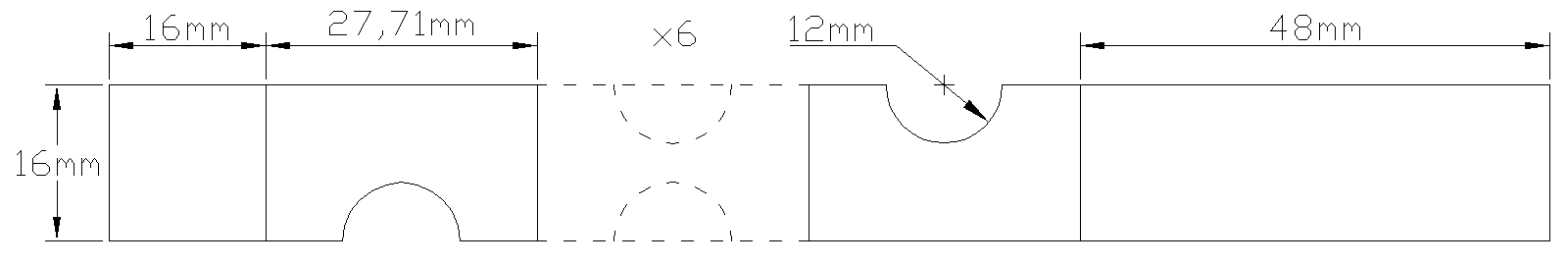}
    \caption{Schematic of the periodic airflow channel unit cell.}
    \label{fig:channel_geometry}
\end{figure}

The numerical simulations were conducted using a representative periodic segment of the airflow channel geometry, following the previously validated methodology proposed in our earlier work on four-row finned-tube heat exchangers \cite{Marcinkowski2022}. In the present study, this approach was adapted to an eight-row configuration, maintaining the geometric and thermal consistency with the original model (Fig. \ref{fig:channel_geometry}).

The segment analyzed represents a single flow channel confined between two adjacent fins and bounded by eight-tube segments arranged in a staggered layout. The geometry includes an upstream buffer zone, repeating tube–fin sections, and a downstream buffer zone that mimics the full airflow path through the exchanger. Key geometric parameters were preserved from the reference model to ensure continuity and reliability of the results. Only minor modifications were introduced to account for the greater number of rows (Fig.~\ref{fig:channel_geometry}). Key dimensions are summarized in Tab. \ref{tab:geomTab}. The effective channel width used in the CFD model is $0.5\,s - 0.5\, \delta_f = 1.43\,\mathrm{mm}$; the model height is $0.5\,p_t = 16\,\mathrm{mm}$; per-row length is $L_r = p_l$.

\begin{table}[h]
\centering
\begin{tabular}{@{}lcl@{}}
\toprule
\textbf{Description}             & \textbf{Symbol} & \textbf{Value} \\ \midrule
Rows {[}-{]}                     & $R$             & 8              \\
Transverse tube pitch {[}mm{]}   & $p_t$           & 32             \\
Longitudinal tube pitch {[}mm{]} & $p_l$           & 27.71          \\
Tube outer diameter {[}mm{]}     & $d_o$           & 12             \\
Tube wall thickness {[}mm{]}     & $\delta_t$      & 0.35           \\
Fin pitch {[}mm{]}               & $s$             & 3              \\
Fin thickness {[}mm{]}           & $\delta_f$      & 0.14           \\
Row length {[}mm{]}              & $L_r$           & 27.71          \\ \bottomrule
\end{tabular}
\caption{Geometric dimensions of a repeatable heat exchanger fragment.}
\label{tab:geomTab}
\end{table}

The main geometric dimensions of the simulated unit cell are as follows: Channel thickness: \SI{1.43}{\milli\meter}, Channel height: \SI{16}{\milli\meter}, Tube diameter: \SI{12}{\milli\meter} (circular cross-section), Fin spacing and tube pitch identical to those in \cite{Marcinkowski2022}, Inlet buffer zone: \SI{16}{\milli\meter}, Repeating tube sections: 8 rows, each spaced by \SI{27.71}{\milli\meter} in flow direction, Outlet buffer zone: \SI{48}{\milli\meter} (Fig.~\ref{fig:channel_geometry}).

\subsection{Boundary conditions and operating points}
At the inlet, a uniform air velocity profile is prescribed with the inlet temperature $T_{in}=20\,^{\circ}\mathrm{C}$. The simulated velocity range covers $1\le w_0 \le 10\,\mathrm{m/s}$, where $w_0$ is air velocity in front of the heat exchanger. The outlet is treated as an \emph{opening} boundary condition and mass-averaged temperature over the outlet section in this solution is monitored. Temperature-dependent air properties are used. This work uses \textbf{only Method 1 (M1)}: isothermal fins and outer walls of the tube at \SI{70}{\celsius}. Previous sensitivity checks over \SI{60} - \SI{80}{\celsius} indicated variations below $1.6 \%$ relative to the \SI{70}{\celsius} baseline; details are provided in  \cite{Marcinkowski2022}.

The fin walls and the outer surfaces of the tubes were maintained at a constant temperature of \SI{70}{\celsius}. Sensitivity checks showed that setting a different constant temperature between \SI{60}{\celsius} and \SI{80}{\celsius} changed the results by less than 2\% from the reference value of \SI{70}{\celsius}. These boundary conditions are widely applicable: e.g., to water–air heat exchangers with liquid–gas heat exchange, as well as evaporators or condensers in which one side is isothermal due to a phase change and the other side is subject to a variable liquid/gas temperature—typical of heat pumps and air-conditioning systems \cite{Marcinkowski2022}.

\subsection{Turbulence model and numerics}
Air-side flow and heat transfer are computed in 3D, steady state, by solving the RANS equations of mass, momentum, and energy with the $k{-}\omega$ Shear Stress Transport (SST) model. The convergence criterion are residuals below $10^{-4}$ for continuity and $10^{-6}$ for momentum and energy. Simulations are performed in Ansys Fluent 2024 R2. Near-wall treatment is fully resolved with $y^+\approx 1$. The key settings are summarized in Tab. \ref{tab:nearwall_inlet_bc}.

\begin{table}[h]
\centering
\begin{tabular}{@{}lc@{}}
\toprule
\textbf{Parameter}          & \textbf{Setting}     \\ \midrule
First-layer thickness       & $0.023\,\mathrm{mm}$ \\
Number of prism layers      & 8                \\
Layer growth rate           & 1.2                  \\
Target near-wall resolution & $y^+\approx 1$       \\
\bottomrule
\end{tabular}
\caption{Parameters of the simulated boundary layer.}
\label{tab:nearwall_inlet_bc}
\end{table}

\section{Numerical Methodology for Row-Specific Heat Transfer Coefficient Determination}
The methodology employed for determining the row-specific heat transfer coefficients (HTC) in this study is derived from the previously validated approach presented in our earlier work \cite{Marcinkowski2024}. The numerical model was adapted for an 8-row fin–tube heat exchanger with staggered tube arrangement and continuous plate fins, and it follows the same assumptions and general computational framework described in the referenced study.

A constant temperature was imposed on the external surfaces of both the fins and the tubes. This simplification allows for a direct determination of the air-side convective HTC without the need for coupled fluid–solid heat transfer modeling. Such an approach is suitable for capturing relative trends in local heat transfer performance and is supported by previous validation efforts showing qualitative and semi-quantitative agreement with experimental measurements \cite{Marcinkowski2024}.

The simulations were performed using ANSYS Fluent 2024 R2. The computational procedure involved the following steps: solving steady-state, single-phase heat transfer simulations under turbulent flow conditions with isothermal walls (no conjugate heat transfer); extracting local mass-average air temperatures behind each tube row from the CFD results; post-processing of thermal and flow field data in Matlab to determine local Reynolds and Nusselt numbers; fitting the row-specific Nusselt number correlations in the form of power-law relationships:
\begin{equation}
\text{Nu}_{a} = x_1 \cdot \text{Re}_{a}^{x_2} \cdot \text{Pr}_a^{1/3}
\end{equation}

The correlation coefficients $x_1$ and $x_2$ were obtained using the least-squares method. The air-side Reynolds number was defined based on the local maximum velocity between adjacent tubes and the hydraulic diameter calculated according to Kays and London \cite{KaysLondon2018}. The hydraulic diameter and fin geometry were held constant throughout the analysis, in line with the baseline configuration established in the earlier study \cite{Marcinkowski2024}.

This methodology enables the identification of row-specific thermal behavior across the exchanger depth and offers a computationally efficient framework for deriving Nusselt number correlations without requiring access to detailed local experimental data.

\section{Mesh independence study}
To ensure the accuracy of the CFD results while minimizing computational costs, a mesh independence study was performed. This procedure followed the validated methodology previously described in our earlier study on row-specific heat transfer in finned-tube heat exchangers \cite{Marcinkowski2024}. As in that work, the Nusselt number for the last (8th) tube row was selected as the evaluation criterion. Due to the cumulative development of thermal and flow fields through upstream rows, the last row requires the most stabilization time and is thus the most sensitive indicator of numerical resolution.

\begin{table}[ht]
    \centering
    \caption{Nusselt number for different mesh resolutions (evaluated at row~8). Difference is relative to the finest mesh (16.3M).}
    \label{tab:Nu_per_elements}
    \begin{tabular}{c c c}
    \toprule
    \multicolumn{1}{c}{\textbf{Elements [M]}} & 
    \multicolumn{1}{c}{\textbf{Nu R8}} & 
    \multicolumn{1}{c}{\textbf{Difference}} \\
    \midrule
    1.1 & 27.20 & 9.36 \% \\
    2.0 & 27.27 & 9.64 \% \\
    4.3 & 25.68 & 3.27 \% \\
    8.2 & 24.99 & 0.47 \% \\
    16.3 & 24.87 & -- \\
    \bottomrule
    \end{tabular}
\end{table}

\begin{figure}[H]
    \centering
    \includegraphics[width=0.7\linewidth]{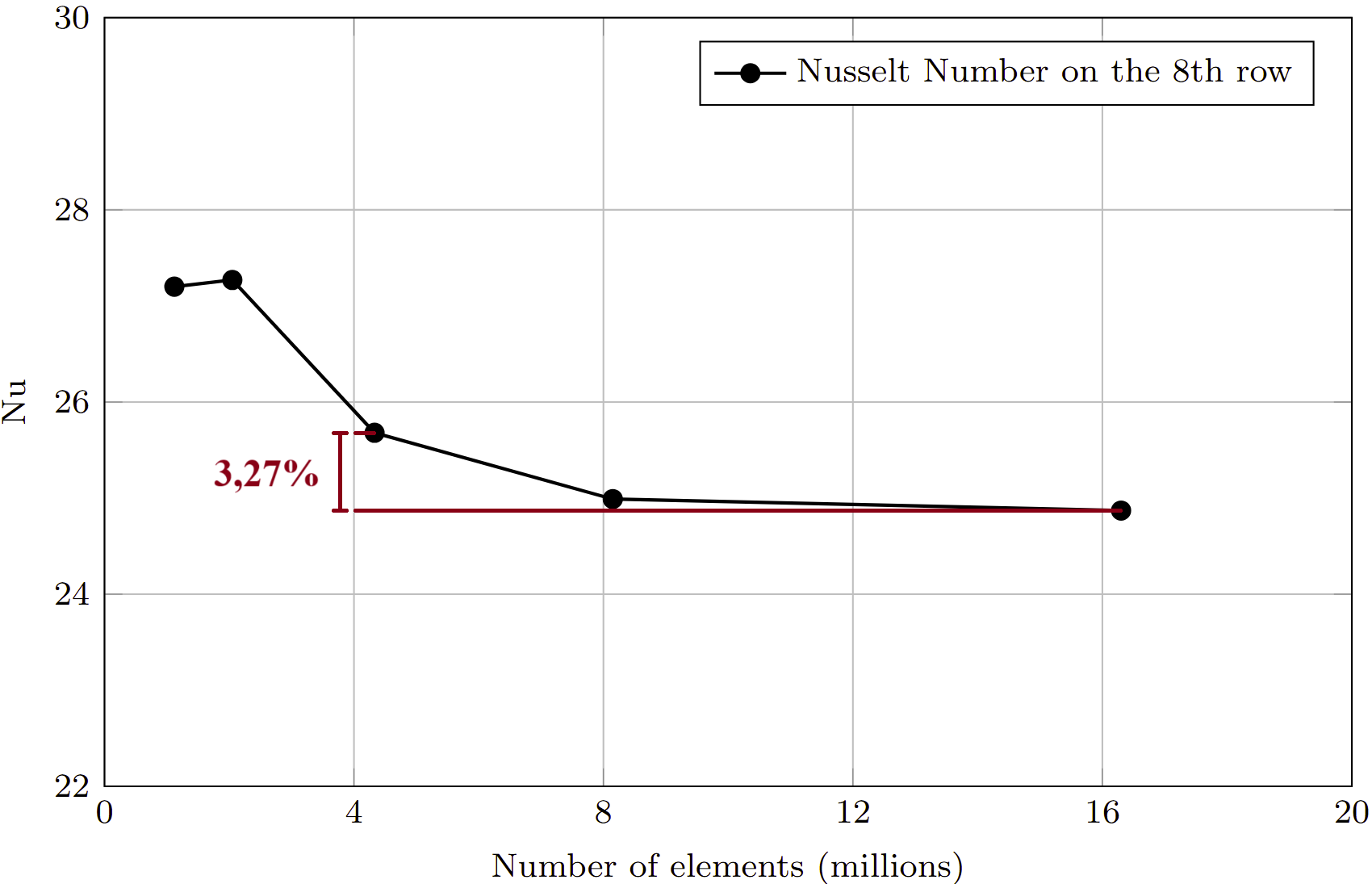}
    \caption{Variation of Nusselt number with mesh refinement for the 8th row. A decrease in Nu is observed as mesh resolution improves, indicating finer boundary layer capture.}
    \label{fig:nu_vs_elements}
\end{figure}

The simulations were conducted using five different mesh densities, ranging from approximately 1.1 to 16.3 million elements. The corresponding Nusselt numbers on the 8th row ($\text{Nu}_{\mathrm{R8}}$) are summarized in Table~\ref{tab:Nu_per_elements} and visualized in Figure~\ref{fig:nu_vs_elements}. The results show a moderate variation in calculated $\text{Nu}_{\mathrm{R8}}$ between 1.1M and 2.0M cells (from 27.20 to 27.27), suggesting that the solution has not yet fully converged. At 4.3M cells, the predicted Nusselt number decreases to 25.68, at 8.2M to 24.99, and at 16.3M elements it further reduces slightly to 24.87.

This trend indicates that finer grids result in a modest drop in local heat transfer coefficients, likely due to better resolution of boundary layer gradients and suppression of artificial diffusion. However, the difference between the 4.3M and 16.3M cases is only 3.27\%, suggesting that a mesh with approximately 4 million elements represents a reasonable trade-off between solution accuracy and computational cost. Consequently, this resolution was used for all subsequent simulations.

\section{CFD Simulation Results and Nusselt Number Correlations}
The Nusselt number is a function of the Reynolds number and the Prandtl number. The Reynolds number (Eq.~\ref{eq:reynolds}) is related to the hydraulic diameter, which was calculated using the definition proposed by Kays and London \cite{KaysLondon2018} with \( w_{\text{max}} \) given by Eq.~\ref{eq:velocity}. Maximum air velocity \( w_{\text{max}} \) was calculated for the minimum airflow cross-section between tubes, and it can exist in different locations depending on FTHE construction. Typically, this hydraulic diameter in FTHE is slightly smaller than double the distance between two adjacent fins. The hydraulic diameter for the analyzed FTHE is \( d_h = 5.35\,\text{mm} \), while the doubled distance between the fins is:

\[
2s - d_f = 5.72\,\text{mm}
\]

The Reynolds number \( \text{Re}_{a,k} \) and maximum velocity \( w_{\text{max},k} \) were calculated for each row separately due to different maximum air velocities caused by varying air temperatures in each tube row.

\begin{equation}
\text{Re}_{a,k} = \frac{w_{\text{max},k} \cdot d_h}{\nu_a}
\label{eq:reynolds}
\end{equation}

Where in Equation~\eqref{eq:reynolds}: \( \text{Re}_{a,k} \) means Reynolds number based on hydraulic diameter (\( d_h \)), \( \nu_a \) denotes air kinematic viscosity.

The maximum velocity (Eq.~\ref{eq:velocity}) in the narrowest cross-section between adjacent tubes is calculated as:

\begin{equation}
w_{\text{max},k} = \frac{(s \cdot p_t)}{(s - d_f)(p_t - d_o)} \cdot \frac{T_{a,i}}{T_{a,0}} \cdot w_0
\label{eq:velocity}
\end{equation}

Where \( s \) means fin pitch, \( p_t \) is transversal fin pitch, \( d_f \) denotes fin thickness, \( d_o \) - outer tube diameter, \( T_{a,i} \) indicates mean mass average air temperature over \( i \)-th tube row, \( T_{a,0} \) is mean mass average air temperature upstream of FTHE.

The unknown parameters \( x_1 \) and \( x_2 \) in the approximation function were determined using least-squares method. The Nusselt number for the \( i \)-th tube row was approximated as:

\begin{equation}
\text{Nu}_{a,k} = x_1 \cdot \text{Re}_{a,k}^{x_2} \cdot \text{Pr}_a^{1/3}
\label{eq:nusselt}
\end{equation}

with constraints:
\[
1300 \leq \text{Re}_{a,k} \leq 5900,\quad \text{Pr}_a = 0.7
\]

This correlation follows from the Colburn analogy \( \text{Nu}/(\text{Re}\cdot\text{Pr}^{1/3}) = f(\text{Re}) \), where \( f(\text{Re}) \) is determined experimentally. The values of the Nusselt number calculated for a given Reynolds number by Eq.~\ref{eq:nusselt} differ by \( \pm \sigma \), where the symbol \( \sigma \) denotes the mean standard deviation of the Nusselt numbers obtained by the CFD modeling.

\begin{figure} [H]
    \centering
    \includegraphics[width=1\linewidth]{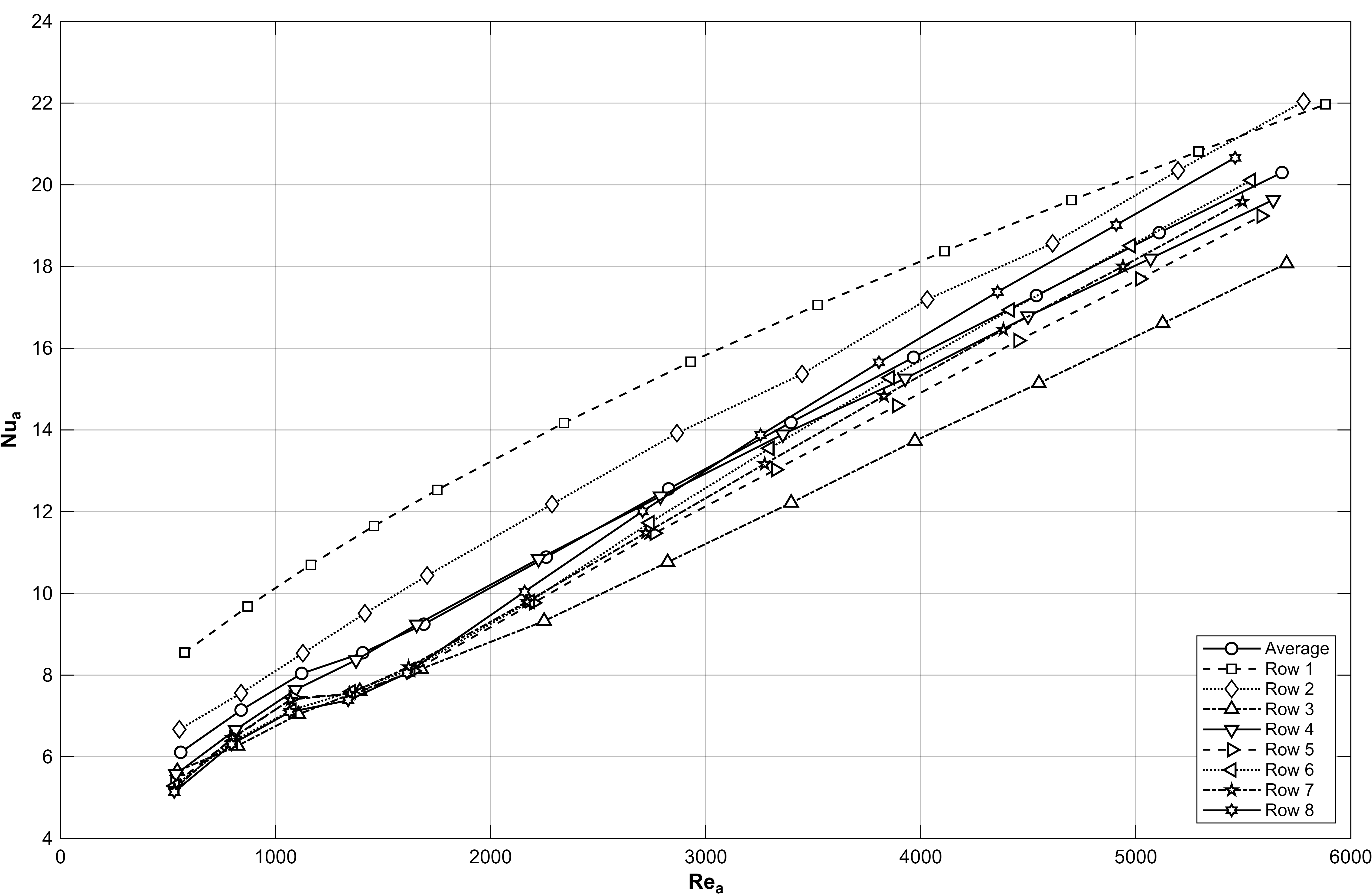}
    \caption{Nusselt number in relation to the Reynolds number for individual tube rows and for the average value for the entire exchanger.}
    \label{fig:8RowNuGraph}
\end{figure}

In the analysis of heat transfer efficiency (Fig.~\ref{fig:8RowNuGraph}) across different rows of the tube heat exchanger, it was observed that the first and second rows exhibit the highest heat transfer performance up to a Reynolds number (Re) of approximately 3400 ($\sim$\SI{6}{\meter\per\second}). Beyond this threshold, the efficiency of these rows is surpassed by the last row. Notably, the second-to-last row does not demonstrate increased efficiency relative to the other rows. Furthermore, at around Re 4200 ($\sim$\SI{7}{\meter\per\second}), the second row surpasses the first in heat transfer efficiency. Rows five through eight experience a significant decline in efficiency within the Re range of 1000 to 1500. Conversely, the heat transfer coefficient in the last row begins to increase more sharply starting at Re approximately 1600 ($\sim$\SI{2.5}{\meter\per\second}). While rows three to seven maintain below-average heat transfer coefficients, rows one, two, and eight consistently perform above the average. Under conditions of low turbulence, rows seven and eight are the least efficient, whereas at high turbulence levels ($\mathrm{Re} > 4000$), rows three and five exhibit the lowest efficiency. Similarities between the increased Nusselt number values for the first, second, and last rows were shown for different numbers of heat exchanger rows in the studies by Marcinkowski et al. \cite{Marcinkowski2024} and Che et al. \cite{Che2022}. These findings highlight the variation in heat transfer behavior across rows and suggest potential opportunities for optimizing heat exchanger design based on its performance.

\begin{figure}[H]
    \centering
    \includegraphics[width=0.5\linewidth]{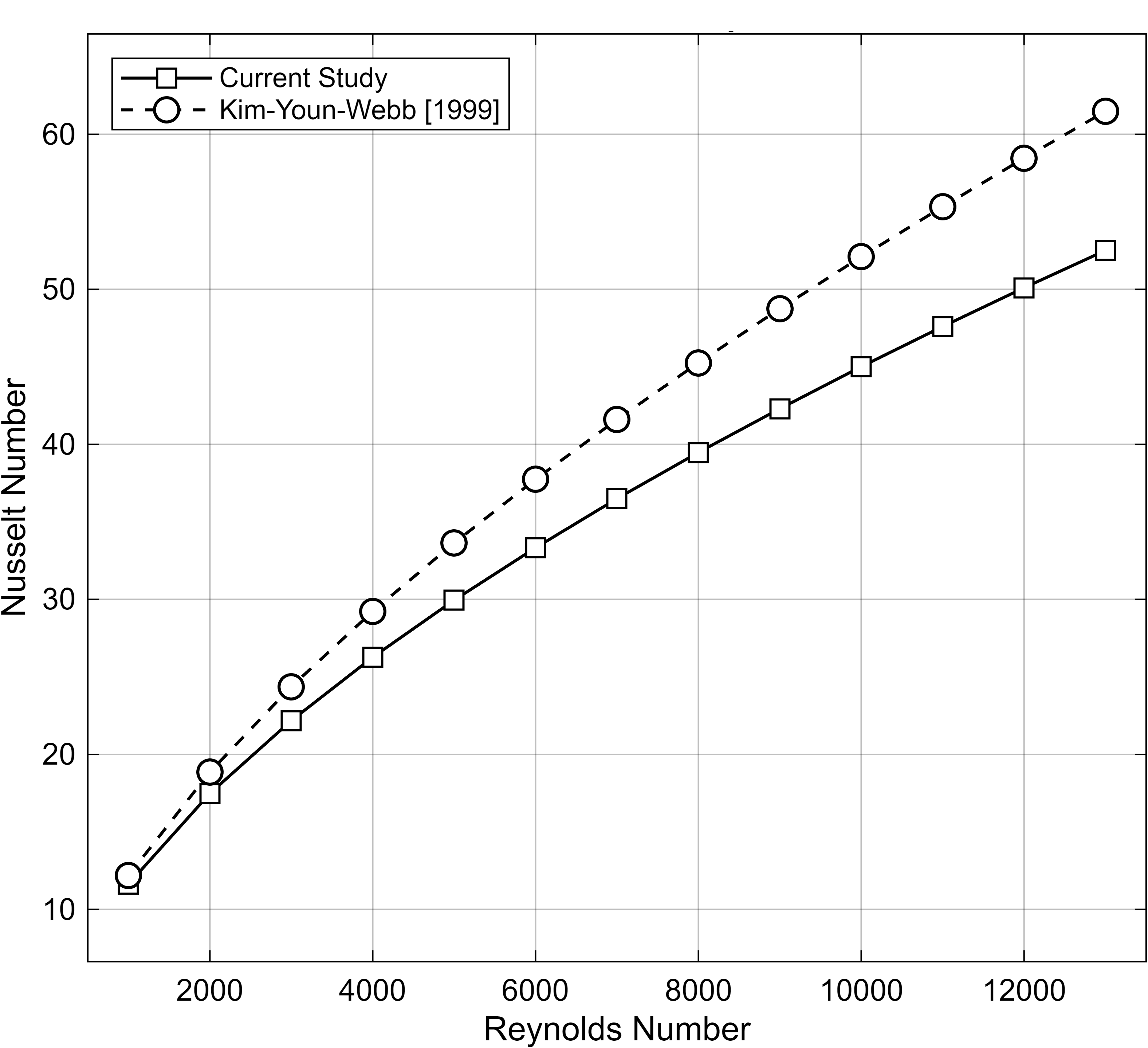}
    \caption{Comparison of current study and results from the literature}
    \label{fig:Kim_comparison}
\end{figure}

The present study's Nusselt number ($Nu$) predictions align closely with those of Kim, Youn, and Webb (1999) \cite{Kim_Youn_Webb} at lower Reynolds numbers ($Re \approx 1000$). However, as $Re$ increases, the results diverge, with a difference of up to 17\% in $Nu$ at $Re \approx 13{,}000$.\footnote{Note: this comparison at $Re \approx 13{,}000$ exceeds the fitted range $1300 \le \mathrm{Re} \le 5900$ and is therefore qualitative.} Kim–Youn–Webb's correlations were derived from experimental data with varying fin-and-tube geometries, which the authors adjusted for analytically. In contrast, the current study employs computational fluid dynamics (CFD), which may contribute to the observed discrepancies at higher $Re$. The experimental data used by Kim–Youn–Webb primarily consisted of heat exchangers with 2--4 tube rows, with only a limited number of 6-row configurations and a single 8-row case. Notably, the 8-row heat exchanger in their dataset had different geometric dimensions compared to the one examined here, potentially influencing the comparison.

\section{Comparison of results with a 4-row exchanger}
\begin{figure}[htbp]
    \centering
    \caption{Comparison of Nusselt numbers versus Reynolds numbers for different tube rows in 4-row and 8-row heat exchangers}
    \label{fig:all_nu_re_comparisons}
    
    \begin{subfigure}[b]{0.48\textwidth}
        \centering
        \includegraphics[width=\linewidth]{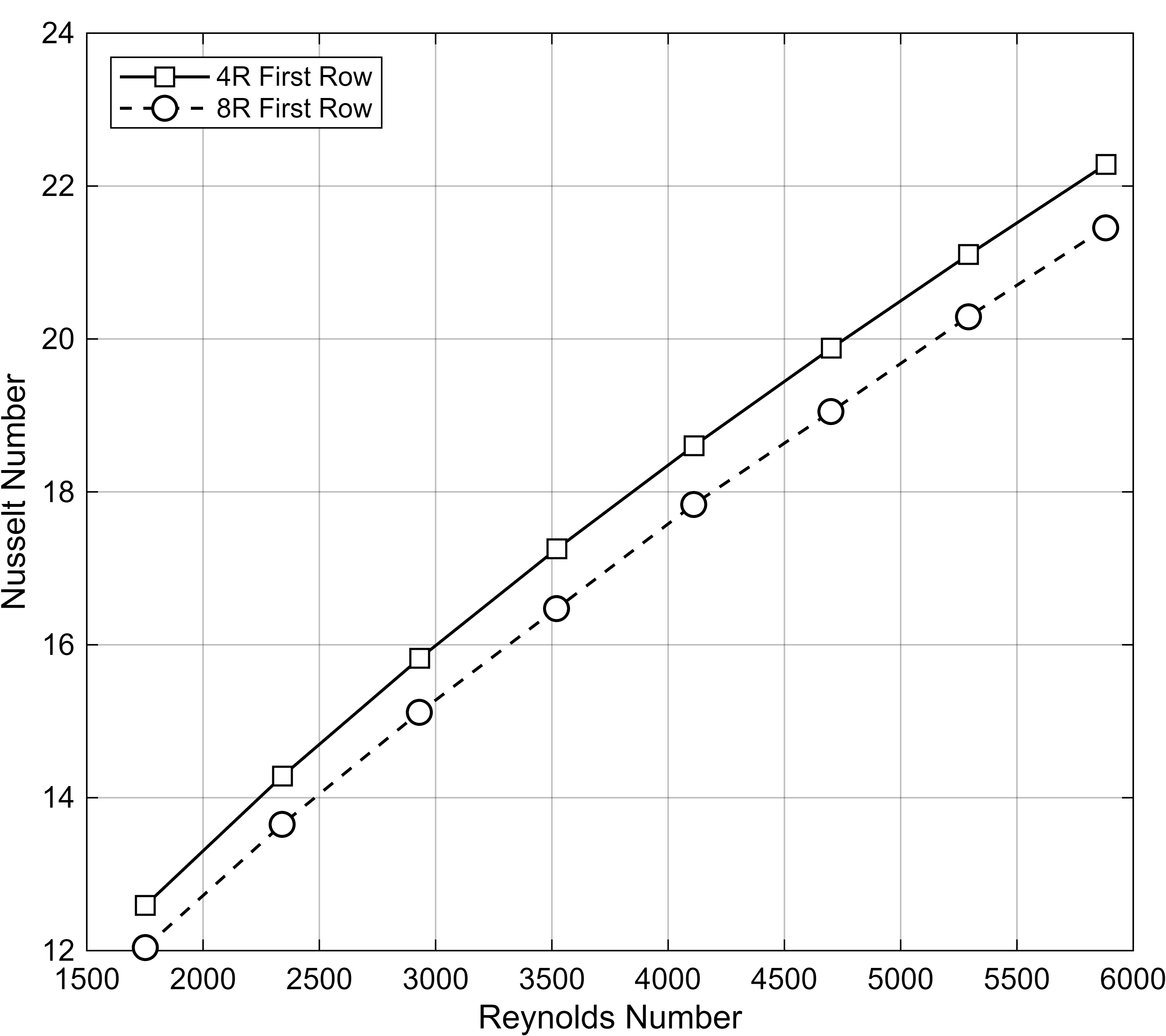}
        \caption{First row}
        \label{fig:first_row_nu_re}
    \end{subfigure}
    \hfill
    \begin{subfigure}[b]{0.48\textwidth}
        \centering
        \includegraphics[width=\linewidth]{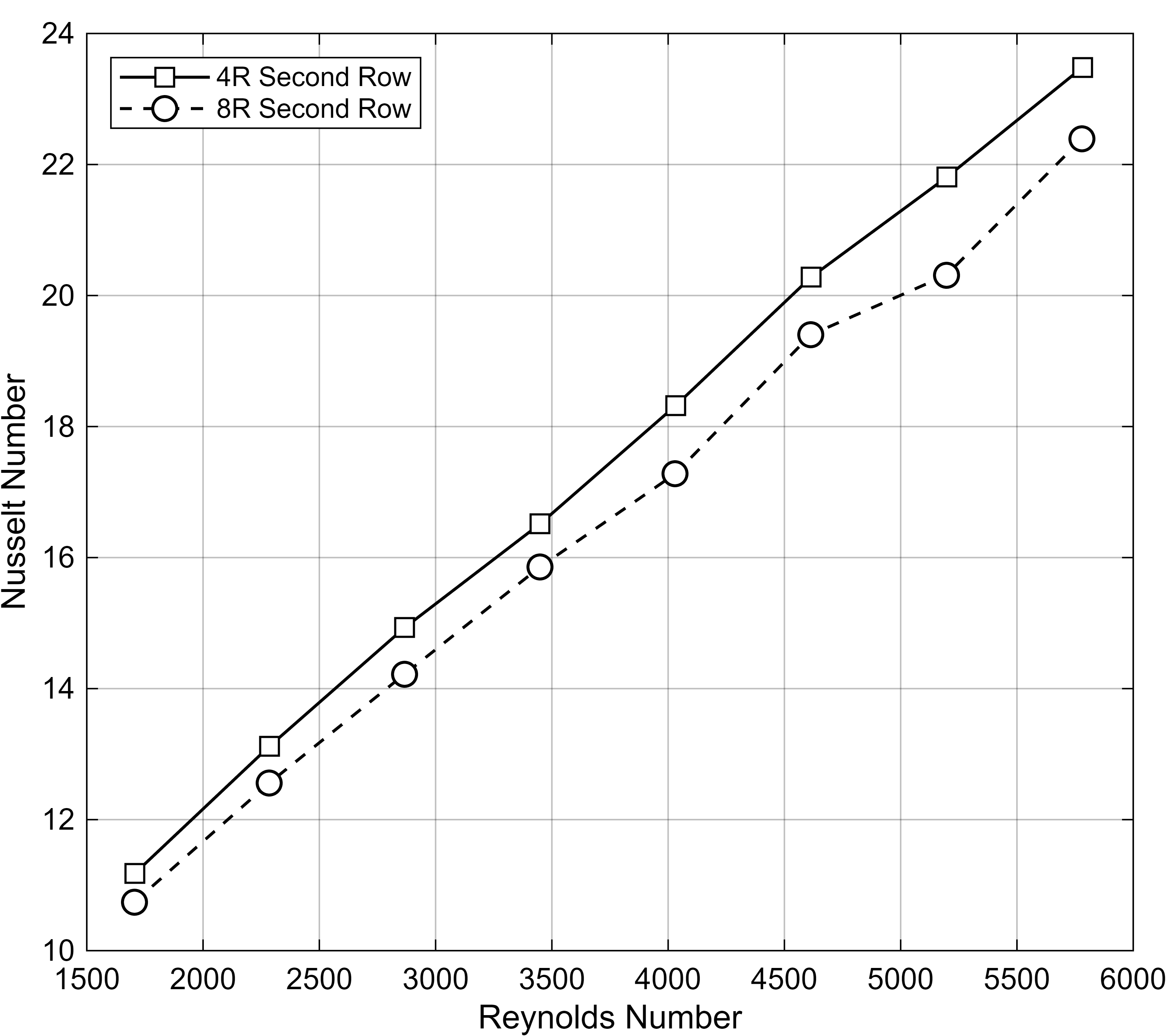}
        \caption{Second row}
        \label{fig:second_row_nu_re}
    \end{subfigure}
    
    \vspace{0.5cm}
    
    \begin{subfigure}[b]{0.48\textwidth}
        \centering
        \includegraphics[width=\linewidth]{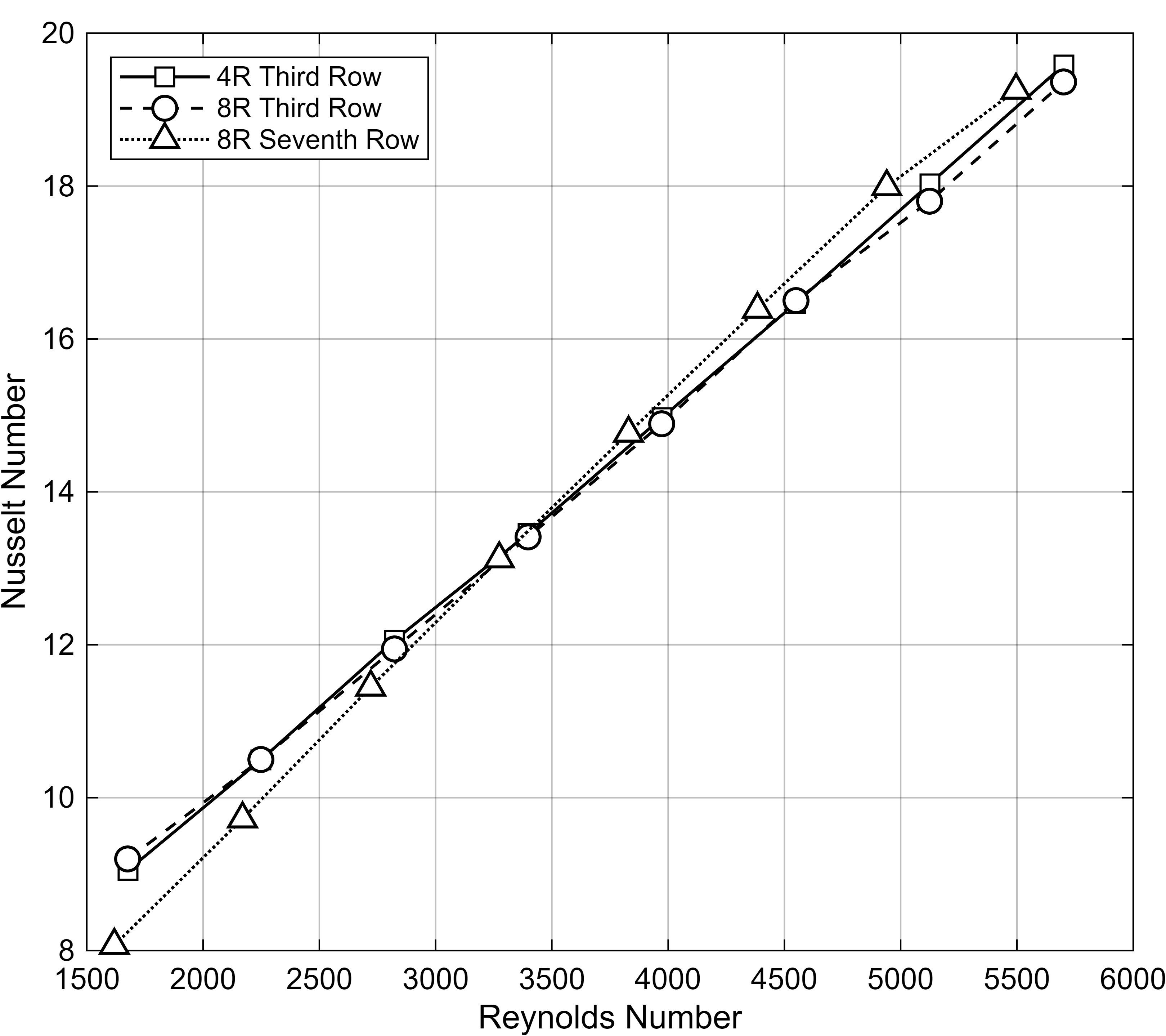}
        \caption{Third rows and 2nd to last row}
        \label{fig:third_row_nu_re}
    \end{subfigure}
    \hfill
    \begin{subfigure}[b]{0.48\textwidth}
        \centering
        \includegraphics[width=\linewidth]{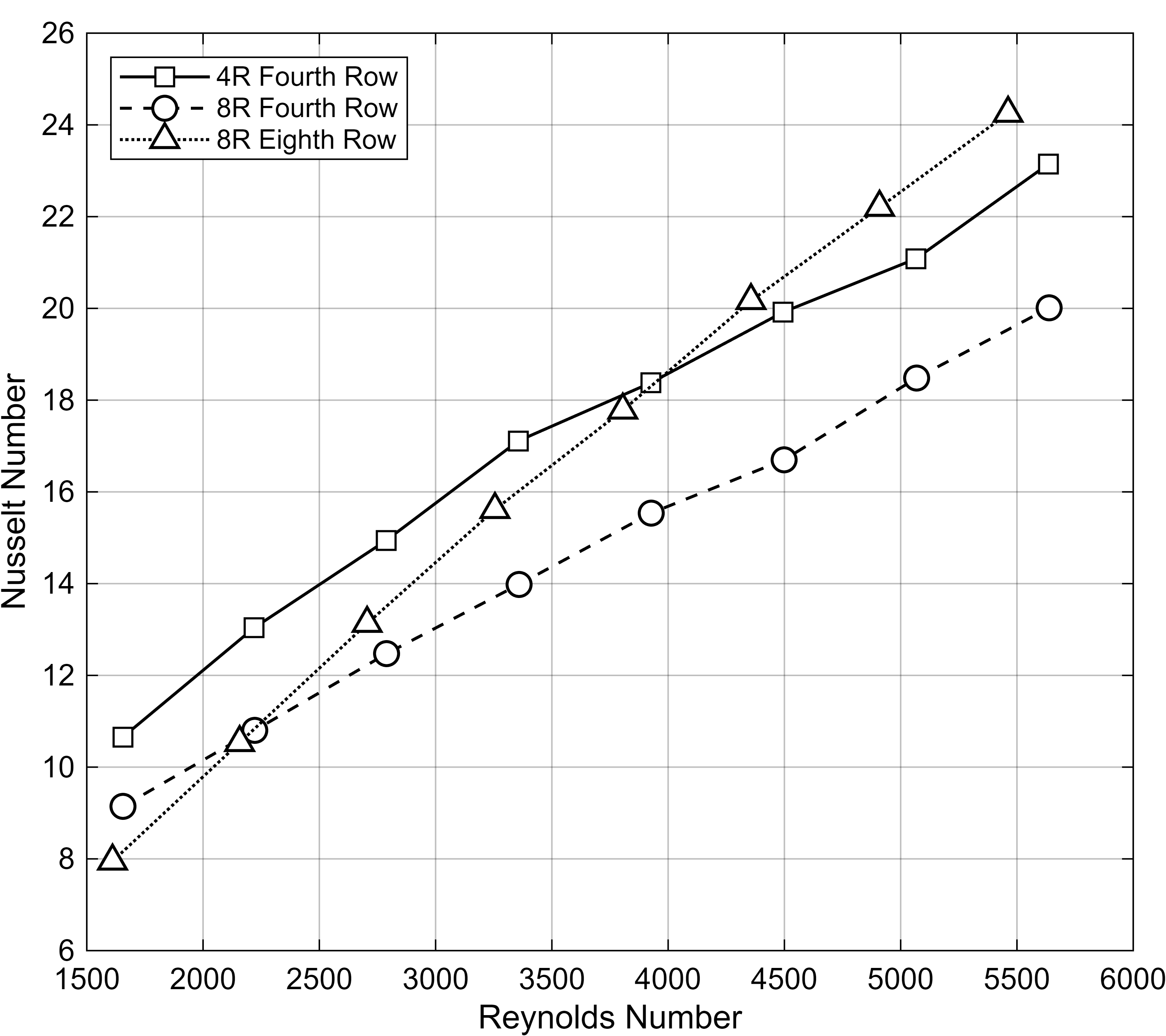}
        \caption{Fourth rows and the last row}
        \label{fig:last_rows_nu_re}
    \end{subfigure}
\end{figure}

Nu for 8R1 (8R1 meaning the 1st row of 8 row exchanger) is lower than 4R1 \hyperref[fig:first_row_nu_re]{(Fig. 4a)}, we would expect them to be the same, but it's unknown whether detachment of the air stream occurs in the same manner in both cases, and whether turbulent eddies take the same amount of space. Those small differences may amount to the slight difference in heat exchange between those rows. Other than an average lower value, other differences between the exchangers are visible as soon as the 2nd row \hyperref[fig:second_row_nu_re]{(Fig. 4b)}, where there is a noticeable dip in how quickly Nu rises for 8R2 between 4500 and 5200 Re.

Third rows \hyperref[fig:third_row_nu_re]{(Fig. 4c)} appear to have a very similar performance, though considering the initial lower Nu of the 8 row exchanger on its first row, it seems its third row's Nu has relatively grown higher. If we consider 4R3 as a second to last row instead, and compare with 8R7, then there is a visible difference in how 8R7 starts at a lower value, but rises more quickly and becomes greater at around 3500 Re, then is slightly higher for Re upwards of 4500.

Comparison of the 4th rows and the 8th \hyperref[fig:last_rows_nu_re]{(Fig. 4d)} show the biggest difference in Nu. 4R4 has a significantly higher Nu than 8R4, but a similar rising trend, while 8R8 behaves similarly to 8R7, meaning it has a lower initial Nu value, but rises more quickly and eventually grows higher than both other rows, albeit at different points, only around 2200 for 8R4, and 4000 for 4R4.

\newpage
\section{Analysis of eddies and dead zones}
\begin{figure}[H]
    \centering
    \includegraphics[width=0.85\linewidth]{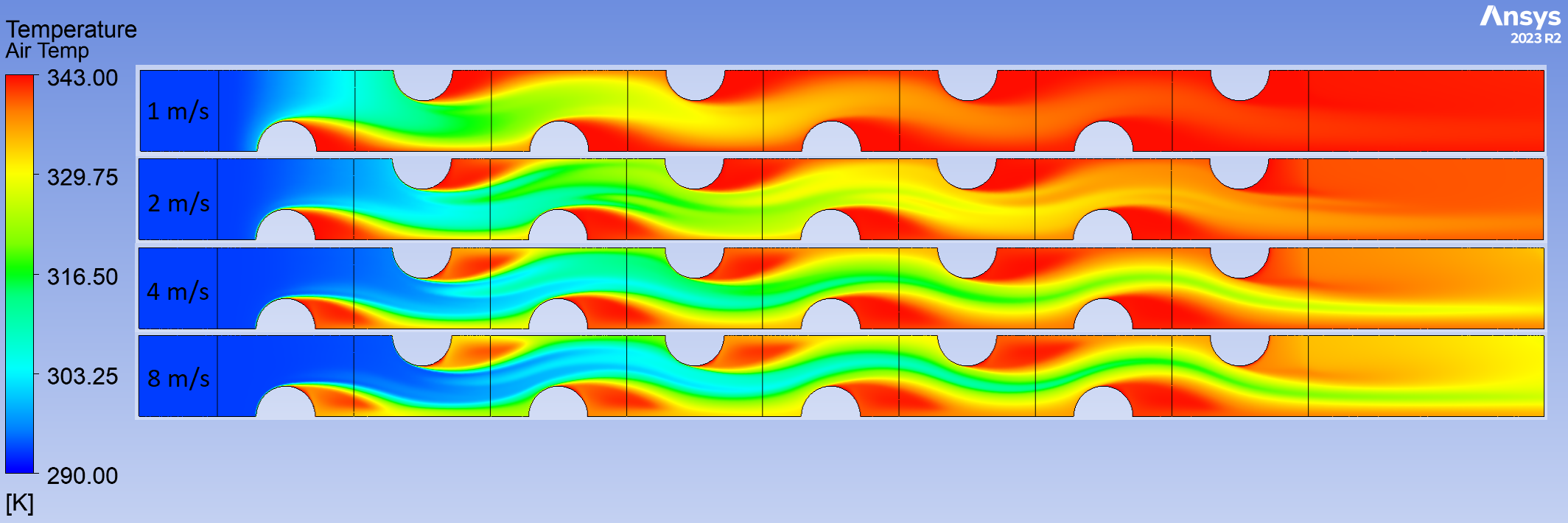}
    \caption{Temperature gradient}
    \label{fig:TempVel}
\end{figure}

\begin{figure}[H]
    \centering
    \includegraphics[width=0.85\linewidth]{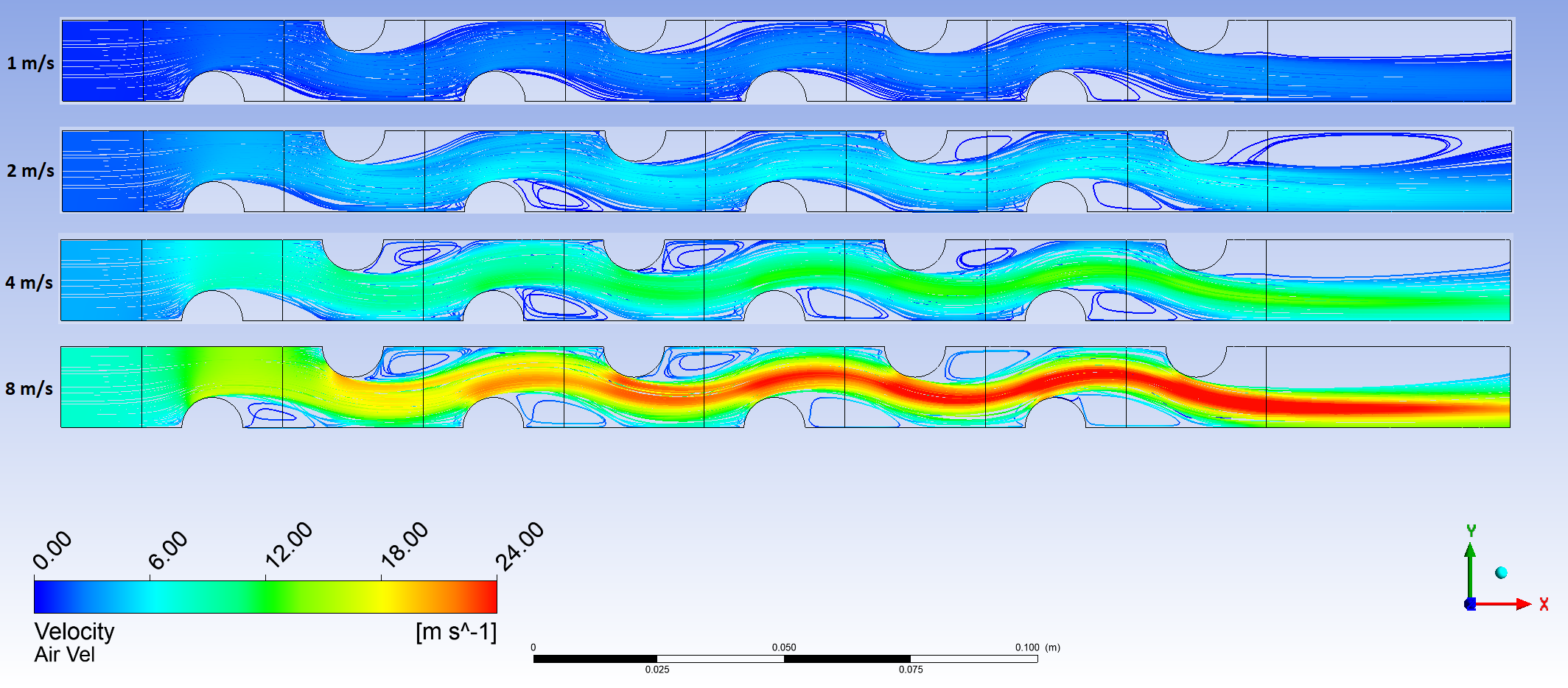}
    \caption{Velocity gradient}
    \label{fig:VelVel}
\end{figure}

The maximum velocity per row exceeds three times the initial air speed, attributable to flow constriction and thermal expansion altering the air-volume-to-space ratio. Kármán vortices develop downstream of the final row \cite{Marcinkowski2024}. At \SI{8}{\meter\per\second} inflow, three thermal layers form immediately behind tubes: (A) a very thin layer at tube temperature, (B) a thin cooler layer ($\sim$\SI{6}{\kelvin} below tube temperature), and (C) an extended layer near tube temperature. This suggests radial airflow from the tube's rear mid-region outward, with reverse flow toward the tube in the central dead zone. Observed in 4-row exchanger animations and validated via full-height CFD models, this phenomenon eliminates symmetry-boundary artifacts \cite[Fig. 18]{Marcinkowski2024}.

\begin{figure}[H]
    \centering
    \includegraphics[width=\linewidth]{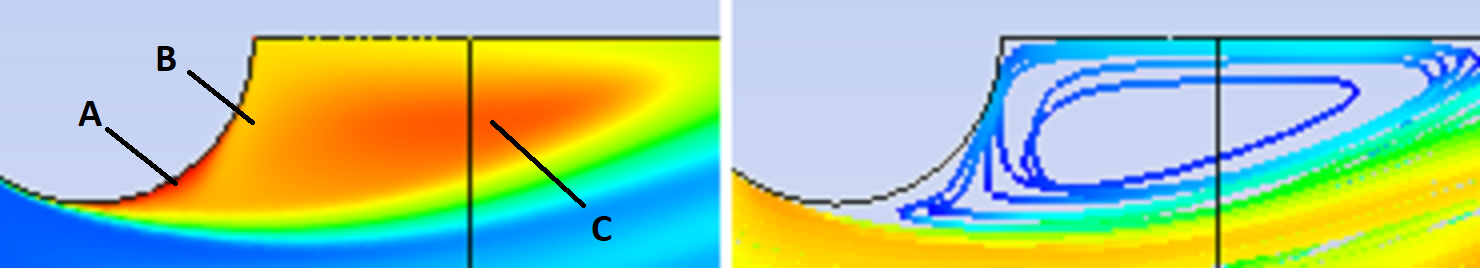}
    \caption{Temperature and velocity distribution at the 2nd row for \SI{8}{\meter\per\second} inflow}
    \label{fig:thermal_layers}
\end{figure}

\section{Discussion}

The observed differences in heat transfer intensity between individual rows result from local flow phenomena that develop as air travels through the tube stack. Three mechanisms influencing the Nusselt number distribution along the depth of the exchanger are particularly pronounced. In the first row, increased heat transfer efficiency is observed, which is related to the presence of a developing boundary layer and the uniform flow rate, which prevents the formation of stagnation zones in the front part of the tubes. The absence of flow disturbances results in uniform scouring of the tube surface and an increased local Nusselt number, especially for lower and moderate Reynolds values (Re < 3400).

For the middle rows (3–7), a significant decrease in efficiency is observed, resulting from wake interference and the increasing effects of wakes behind the preceding tubes. The flow separates behind each tube, creating low-velocity recirculation zones, visible in the velocity-temperature field diagrams (fig.~\ref{fig:TempVel}, \ref{fig:VelVel}). The developing boundary layer flattens and merges between the rows, leading to a channeling phenomenon—a significant portion of the flow bypasses the active heat transfer surfaces, moving along narrow, low-resistance paths.

In the last row (row 8), for Re > 3400, a secondary increase in heat transfer intensity is noticeable. The absence of another row of tubes allows for the free development of Kármán vorticity and non-stationary recirculation structures. These phenomena, as demonstrated in previous work \cite{Marcinkowski2024}, cause intense dispersion and disruption of the boundary layer, leading to a local increase in the heat transfer coefficient. The characteristic three-layer temperature structure (fig.~\ref{fig:thermal_layers})—with a thin boundary layer, a mixing zone, and an outer advection layer—confirms the presence of dynamic heat transport mechanisms behind the last row.

The change in row efficiency with increasing Reynolds number reflects a transition in flow character—from laminar, viscosity-dominated at low Re to highly turbulent, eddy-dominated at high Re. These results have a direct impact on design optimization. The observed inhomogeneity in thermal efficiency between rows indicates the possibility of reducing their number without loss of efficiency, provided that the selected rows operate within the Reynolds range that ensures maximum efficiency.

\section{Conclusions}

This study demonstrates that the row-wise heat transfer in multi-row plate–fin–and–tube heat exchangers is governed by vortex dynamics that vary along the flow direction. Three phenomena explain the observed unsteady flow Nusselt distribution: (i) a developing boundary layer and clean stagnation at the front row yield the highest HTC at low–moderate $Re$; (ii) in the core (rows 3–7), wake merging and quasi-steady recirculation create low-shear “dead zones,” thickening the thermal boundary layer and depressing Nu; (iii) at the last row, the absence of a downstream obstacle permits free shear-layer roll-up and Kármán-street formation, which re-energize mixing and thin the boundary layer, producing a secondary HTC rise at higher $Re$ (cf. Figs.~\ref{fig:8RowNuGraph}, \ref{fig:thermal_layers}). This mechanism also rationalizes the observed cross-over where rear performance catches up with—and can locally exceed—mid-row values as turbulence intensifies.

We also provide row-resolved air-side correlations
\[
\mathrm{Nu}_{a,k}=x_1\,\mathrm{Re}_{a,k}^{\,x_2}\,\mathrm{Pr}_a^{1/3}\quad (k=1,\dots,8),
\]
valid for $1300\le \mathrm{Re}_{a,k}\le 5900$ and $\mathrm{Pr}_a=0.7$. The coefficients $(x_1,x_2)$ capture the discrete transport regimes: lower slopes for the shielded middle rows and a steeper high-$Re$ response for the last row, consistent with vortex-induced mixing. Within the studied range, the correlations reproduce CFD-derived Nusselt numbers with a mean deviation of $\pm\sigma$ and are suitable for preliminary design and control-oriented models.

Describing the vortex-controlled gain at the exchange outlet allows for selective order reduction without thermal penalty at the appropriate $Re$ window, lowering pressure drop and fan power; and targeted air-side enhancements (e.g., local fin features near the rear row) where they are most effective.

Current CFD approaches with constant wall temperature and stationarity reflect the dominant trends but do not account for the statistics of unsteady vortices. Future work should employ URANS/LES/DNS with vortex-identification diagnostics (e.g., $Q$ or $\lambda_2$) and row-resolved experiments (e.g., color-reaction HTC mapping) across dry/wet conditions to generalize the last-row intensification and extend the correlations beyond the baseline geometry.

\end{document}